IAC-19.A7.3.10x49665

# Linear-mode avalanche photodiode arrays for low-noise near-infrared imaging in space

**James Gilbert[a]\*, Alexey Grigoriev[a], Shanae King[a], Joice Mathew[a], Rob Sharp[a], Annino Vaccarella[a]**

[a] *Research School of Astronomy & Astrophysics, Australian National University, Mt Stromlo Observatory, Cotter Road, Weston Creek, ACT 2611, Australia*
\* Corresponding Author

**Abstract**

Astronomical observations often require the detection of faint signals in the presence of noise, and the near-infrared regime is no exception. In particular, where the application has short exposure time constraints, we are frequently and unavoidably limited by the read noise of a system. A recent and revolutionary development in detector technology is that of linear-mode avalanche photodiode (LmAPD) arrays. By the introduction of a signal multiplication region within the device, effective read noise can be reduced to <0.2 e$^-$, enabling the detection of very small signals at frame rates of up to 1 kHz. This is already impacting ground-based astronomy in high-speed applications such as wavefront sensing and fringe tracking, but has not yet been exploited for scientific space missions. We present the current status of a collaboration with Leonardo MW – creators of the 'SAPHIRA' LmAPD array – as we work towards the first in-orbit demonstration of a SAPHIRA device in 'Emu', a hosted payload on the International Space Station. The Emu mission will fully benefit from the 'noiseless' gains offered by LmAPD technology as it produces a time delay integration photometric sky survey at 1.4 μm, using compact readout electronics developed at the Australian National University. This is just one example of a use case that could not be achieved with conventional infrared sensors.

**Acronyms/Abbreviations**

| | |
|---|---|
| ANU | Australian National University |
| CDS | Correlated double sampling |
| CMOS | Complementary metal-oxide-semiconductor [technology] |
| ESO | European Southern Observatory |
| FPGA | Field-programmable gate array |
| LmAPD | Linear-mode avalanche photodiode |
| MOVPE | Metalorganic vapour phase epitaxy |
| NIR | Near-infrared |
| ROIC | Readout integrated circuit |
| TDI | Time delay integration |
| U | CubeSat volume 'unit' (100 mm cube) |

## 1. Introduction

The past decade has seen the emergence of effectively noise-free near-infrared (NIR) detectors from Leonardo (formerly Selex ES), in the form of the 'SAPHIRA' range of linear-mode avalanche photodiode (LmAPD) arrays. The majority of this work has been funded by the astronomical community, namely the European Southern Observatory (ESO) and the University of Hawaii (UH), to address challenges for future optical telescopes.

*1.1 The SAPHIRA LmAPD array*
The current-generation SAPHIRA detector (Fig. 1) has 320×256 pixels on a 24 μm pitch. It is a HgCdTe device, sensitive to a nominal wavelength range of 0.8–2.5 μm, with a CMOS readout integrated circuit (ROIC) that provides several readout and reset modes. The device is produced using Metal Organic Vapour Phase Epitaxy (MOVPE) technology.

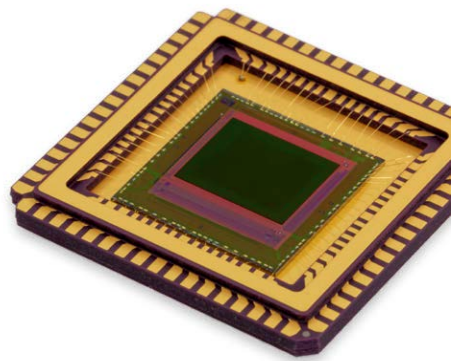

Fig. 1. Image courtesy of Leonardo MW. The SAPHIRA is a 320×256 pixel linear-mode avalanche photodiode array capable of 'noiseless' readouts via an upstream signal multiplication of several hundred.

What is novel about the SAPHIRA is its multiplication region (Fig. 2). This provides linear-mode (as opposed to Geiger-mode) avalanche multiplication of the electrons produced by NIR photons in the absorber layer (Fig. 3), yielding a noiseless signal gain *before* the pixels are read, and hence before the injection of readout noise. Gains of several hundred are possible and can be set with a single bias voltage. This has profound implications for high-speed and/or photon-starved NIR applications, virtually eliminating the effect of readout noise when the avalanche gain is high.

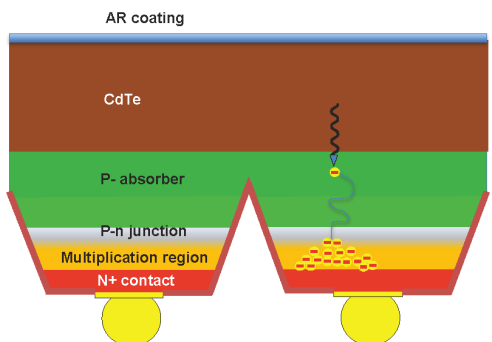

Fig. 2. Image courtesy of Leonardo MW [1]. Current-generation SAPHIRA devices use bandgap engineering and MOVPE technology to achieve linear-mode electron avalanche multiplication of several hundred times, while supressing dark current at high gains.

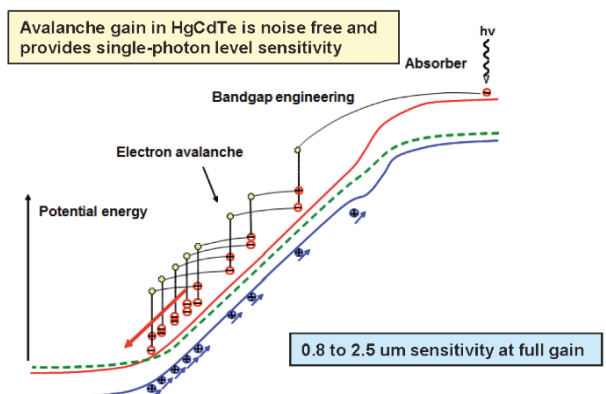

Fig. 3. Image courtesy of Leonardo MW [1]. Photons with wavelengths 0.8–2.5 μm produce electrons in the absorber layer just like a traditional HgCdTe detector, but these electrons are then multiplied within the biased 'multiplication region', boosting signal levels without introducing significant excess noise [2].

*1.2 Enabling new applications in space*

SAPHIRA LmAPD technology enables a variety of use cases not possible with conventional HgCdTe arrays, either through direct use of signal multiplication for detecting faint sources (single photon-counting has been demonstrated [3]), or by 'trading' some or all of this increased sensitivity for speed. The latter has already revolutionized ground-based astronomy by enabling fast wavefront sensing and fringe tracking applications.

Many space-based instruments will also benefit from SAPHIRA technology, especially as array sizes increase (see section 4). One clear application is time delay integration (TDI) imaging without the need to 'point and stare' at objects. Instead, many low-brightness images can be captured at a high frame rate, then aligned and stacked to produce the same result. This is the premise of the ANU 'Emu' mission (section 3), but could just as readily be applied to other photometric or hyperspectral instruments for Earth observation and remote sensing.

## 2. Performance

The majority of performance measurements and characterization of SAPHIRA arrays to date has come from the astronomical community. Continued collaboration between major astronomical groups and Leonardo has already resulted in significant improvements to the original SAPHIRA arrays.

*2.1 Readout noise*

The defining feature of SAPHIRA LmAPD arrays is their 'noiseless' readout capabilities at high avalanche gains. The reduction in noise is an *effective* reduction, in that the noise introduced by the ROIC at time of readout (typically of order ~20 $e^-$) can be divided by the upstream avalanche gain of the signal.

The European Southern Observatory has demonstrated [2] readout noise of <0.2 $e^-$ at avalanche gains in excess of 400. Sub-electron noise values are shown in Fig. 4 for a selection of gains and exposure times. Note that dark current performance has drastically improved since these measurements (see 2.2, below), suppressing noise for longer exposures.

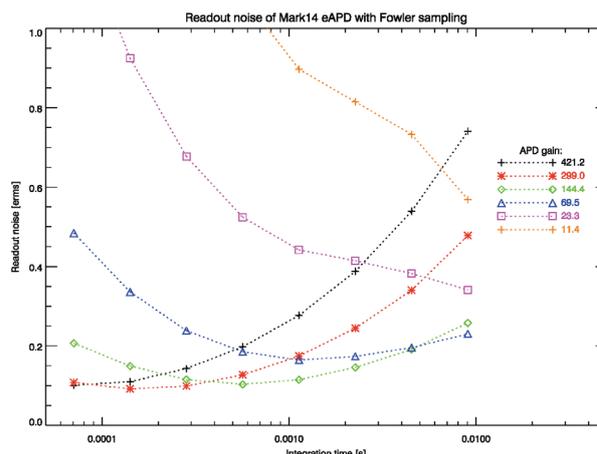

Fig. 4. Image courtesy of ESO [2]. Effective SAPHIRA readout noise values of <0.2 $e^-$ have been demonstrated for high gains. This data used increased frame averaging for longer exposures, hence the reduction in noise for longer exposures at low gains. The increase in noise with exposure time for high gains is due to dark current (since improved).

*2.2 Dark current*

Early versions of the SAPHIRA detector produced significant dark current, restricting their suitability for applications at slower frame rates. The dark current was found to be dominated by a glowing component on the ROIC [4,5], which has been fixed in the latest SAPHIRA devices.

Recent measurements have shown very low dark current values similar to the best infrared detectors. Fig. 5 shows recent dark current data from ESO [2] and the University of Hawaii (UH) [4], and its dependence on detector temperature and avalanche gain.

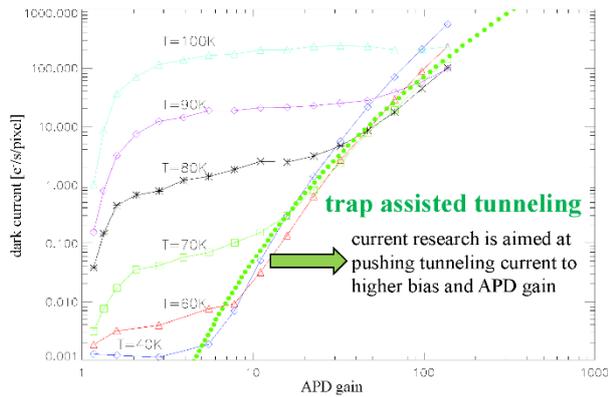

Fig. 5. Image courtesy of Leonardo MW [1]. Recent dark current measurements by ESO [2] and UH [4] for SAPHIRA 'Mk 13/14' devices show very low values at low temperatures and unity gain. The increase in noise with avalanche gain may limit exposure times at high gains, and is the subject of ongoing research.

*2.3 Speed*

The signal multiplication provided by the SAPHIRA's avalanche gain makes these detectors ideal for high-speed applications. Indeed, the early development of SAPHIRA devices was funded for use in high-speed adaptive optics and interferometry instruments for ground-based astronomy [6,7].

The current generation of 320×256 SAPHIRA arrays supports a pixel rate of up to 10 MHz for each of up to 32 parallel readout channels, yielding a maximum theoretical full-frame rate of 3.9 kHz. Windowed (area of interest) readout is also possible, at higher frame rates.

In reality, readout electronics limitations are likely to set the maximum readout speed. This is usually due to analog-to-digital conversion bottlenecks, where high-resolution (e.g. 16-bit) data must be shifted, often serially, for every pixel that is read out.

ANU recently completed an on-sky 'Lucky Imaging' demonstration with a SAPHIRA reading out at ~600 Hz full-frame (~300 Hz CDS) and ~4.4 kHz for a 96×100 window (~2.2 kHz CDS).

The SAPHIRA ROIC supports global and row-by-row reset modes, allowing rolling shutter readouts with minimal lost integration time.

*2.4 Quantum efficiency / sensitivity*

While the nominal wavelength range of the SAPHIRA is 0.8–2.5 μm, the avalanche multiplication region somewhat complicates their total spectral response. This is because photons with wavelength longer than 2.5 μm will produce electrons in the multiplication region itself, penetrating a distance that is wavelength-dependent out to ~3.5 μm.

It follows that the effective quantum efficiency in the range 2.5–3.5 μm depends on the selected avalanche gain, with higher gains diminishing the significance of long-wavelength-induced electrons that do not experience full avalanche multiplication. This is illustrated in Fig. 6, showing measured sensitivity ('quantum efficiency') for unity gain and high gain.

Blocking of light beyond 2.5 μm can be achieved with a cold short-pass filter in the optical path, in tandem with a scientific filter.

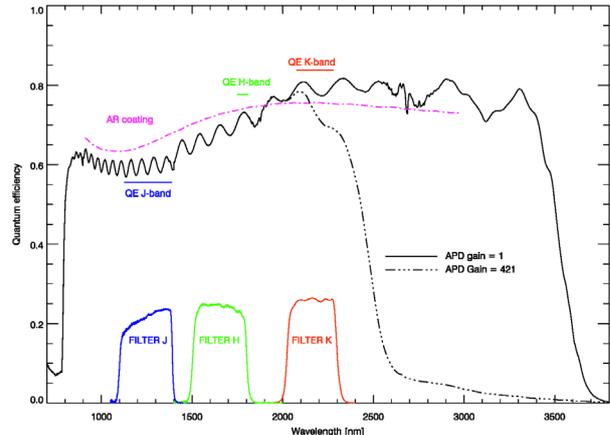

Fig. 6. Image courtesy of ESO [2]. Measured sensitivity ('quantum efficiency') of a SAPHIRA 'Mk 14' detector at unity and high (421×) gain. The response beyond 2.5 μm depends on the avalanche gain, as electrons produced by 2.5–3.5 μm photons experience only a fraction of the avalanche multiplication. The high-gain curve has been normalized.

*2.5 Space-readiness*

As reported by Baker et al. [1], gamma and proton radiation testing has been undertaken on the SAPHIRA Mk 14 device as part of a European Space Agency (ESA) test campaign to establish the radiation hardness of the technology. Neither produced a permanent change to device performance or array defects.

**3. The Emu mission**

ANU aims to conduct the first in-orbit demonstration of SAPHIRA LmAPD technology with 'Emu', a wide-field time delay integration (TDI) imager to be hosted on the International Space Station (ISS) in 2021.

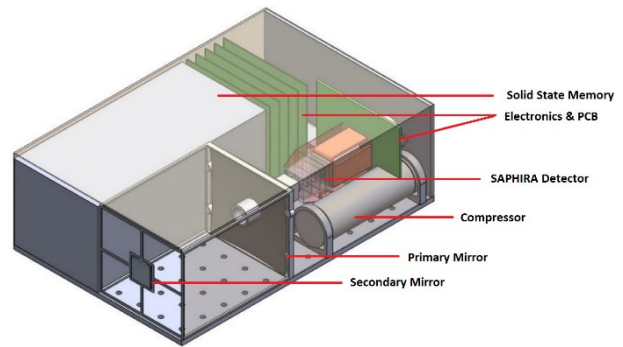

Fig. 7. Conceptual design of the 6U 'Emu' payload, featuring a SAPHIRA LmAPD array and ANU readout electronics. Note that the payload will in fact point 'up' to zenith.

The Emu payload (Fig. 7) has a '6U' format (300 mm × 200 mm × 100 mm) and comprises a small telescope with an ~85 mm square aperture and ~1.2° field of view, feeding a SAPHIRA detector cooled to 80 K by a Thales miniature linear Stirling cycle cryo-cooler. Flight electronics, including readout electronics for the SAPHIRA, are being developed at ANU. Approximately 2U of volume is reserved for high-capacity solid state hard drives for the return of all raw data back to Earth.

*3.1 Science case*

Emu will exploit the 'noiseless' readout capability of the SAPHIRA detector to perform a photometric survey of the sky in the 1.4 µm 'water band' between the astronomical 'J' and 'H' bands, to a depth of $m_{AB} \approx 13$ (H). It will do this using TDI imaging without any active pointing, using only the precessing ISS orbit to sweep the sky. This is an example of a use case that would be impossible using traditional NIR detectors, which would need to point and stare at a field to obtain the same signal-to-noise ratio.

Observing the sky at 1.4 µm – not possible from Earth due to moisture in our atmosphere – will open a unique window into the physics of stellar atmospheres, to determine oxygen abundance in these systems through measurement of their water absorption-bands.

*3.2 Technology demonstration*

A key goal of the Emu mission is to increase the Technology Readiness Level (TRL) of transformative technologies and systems to flight-proven 'TRL 9'. This includes flight testing of both the SAPHIRA LmAPD array (currently TRL 7) and the associated ANU-built readout electronics.

The Emu detector readout electronics system is compact, modular and flexible. It comprises a central 'timing board' based around a Xilinx Kintex Ultrascale field-programmable gate array (FPGA), with fully configurable clock patterns and 32 low-noise analog-to-digital conversion channels. A modular system architecture means that many traditional CMOS detectors are also supported, including Teledyne 'Hawaii' H2RG and H4RG devices.

Emu will operate with a raw frame rate of ~50 Hz (25 Hz CDS) to avoid smearing. This rate is set by Emu's focal plane scale of 10 arcsec/pixel, combined with the orbital period of the ISS. While this speed is rather modest, the design caters for other space-borne applications, including high-resolution Earth observation with frame rates approaching 1 kHz.

The system is designed for low Earth orbit, with triple modular redundancy (TMR) for protection against radiation upset events, and scrubbing techniques for recovery. Qualification for the space environment and launch conditions will be achieved at Australia's National Space Test Facility at ANU (Fig. 8).

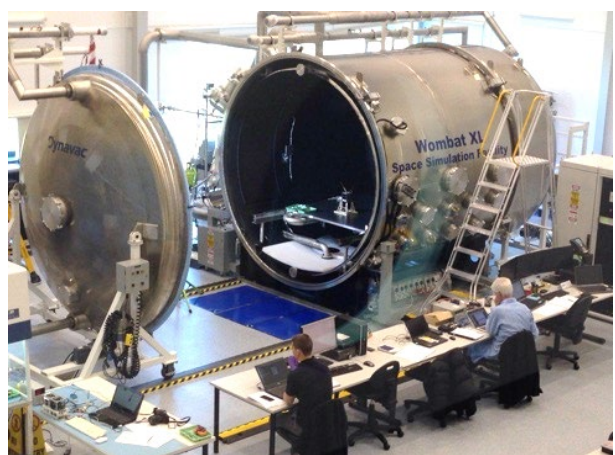
Fig. 8. Australia's National Space Test Facility at ANU provides spacecraft qualification and test services including thermal cycling, thermal vacuum, and launch vibrations.

Prototype electronics (Fig. 9) have been tested in the lab and on-sky at the ANU 2.3 m telescope (Fig. 10). The final board stack will occupy a volume of approximately 0.5 U, comprising four printed circuit boards (PCBs) with flex-PCB interconnects, plus cryo-rated preamplifiers near the focal plane for increased noise immunity [8].

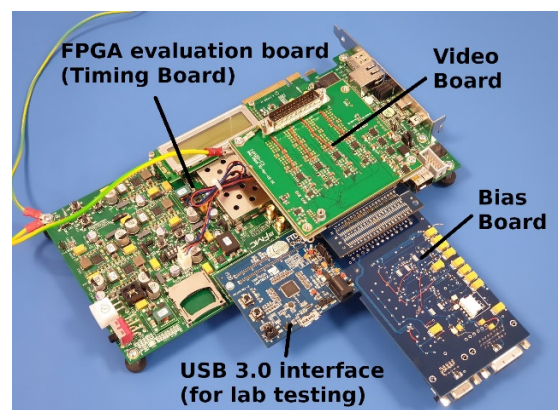
Fig. 9. Prototype Emu readout electronics, including a commercial FPGA evaluation board, one of two 16-channel 'video boards' for analog-to-digital conversion, and a 'bias board' for detector biases and clock conditioning.

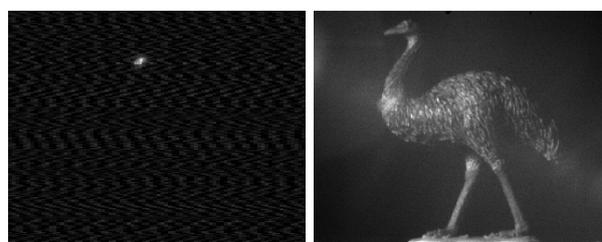
Fig. 10. The prototype readout electronics system in Fig. 9 has been tested with a SAPHIRA detector on-sky (star in H-band, left) for 'Lucky Imaging' at the ANU 2.3 m telescope (Siding Spring Observatory), and recently in the lab (plastic Emu in H-band, right) with improved noise performance.

### 4. Future LmAPD array developments

Leonardo is currently producing a 1k×1k LmAPD array, following development funded by the University of Hawaii and NASA [1]. The device has 15 μm pixels, 16 readout channels, and is designed for low dark current. It offers flexible readout options, including windowing and reference outputs.

ANU has partnered with Leonardo as part of a UK Centre for Earth Observation Instrumentation (CEOI) program to characterize larger format arrays in a simulated space environment (without radiation), providing early-stage TRL increase.

A natural evolution of a 1k×1k device is a 2k×2k device. This would unlock a key area in focal plane technologies for space- and ground-based instrumentation alike.

### 5. Conclusions and prospects

Focused collaborative efforts between Leonardo MW and the ground-based astronomical community have yielded a paradigm-changing technology in the form of the SAPHIRA detector, a HgCdTe LmAPD array offering sub-electron effective read noise.

Current-generation SAPHIRA devices have demonstrated useable avalanche gains of several hundred and dark currents similar to the best conventional infrared sensors on the market.

The 'Emu' mission will flight-qualify an integrated detector and cooler assembly (IDCA) based around the existing SAPHIRA array, with compact FPGA-based readout electronics developed by ANU. Launch is currently expected in 2021.

Current development efforts at Leonardo aim to further reduce dark current while increasing array size to 1k×1k and beyond, paving the way for the next generation of space-borne NIR instruments.

### Acknowledgements

The author wishes to thank Leonardo MW Ltd., the European Southern Observatory, and the University of Hawaii, for their permission to reproduce data and results in this paper.

### References

[1] I. M. Baker et al., Linear-mode avalanche photodiode arrays in HgCdTe at Leonardo, UK: the current status, in Image Sensing Technologies: Materials, Devices, Systems, and Applications VI, 2019, vol. 10980, no. May, p. 20.

[2] G. Finger et al., Sub-electron read noise and millisecond full-frame readout with the near infrared eAPD array SAPHIRA, in Adaptive Optics Systems V, 2016, vol. 9909, no. July 2016, p. 990912.

[3] D. Atkinson, D. Hall, S. Jacobson, and I. M. Baker, Photon-counting Properties of SAPHIRA APD Arrays, Astron. J., vol. 155, no. 5, p. 220, 2018.

[4] D. E. Atkinson, D. N. B. Hall, S. M. Jacobson, and I. M. Baker, Dark Current in the SAPHIRA Series of APD Arrays, Astron. J., vol. 154, no. 6, p. 265, 2017.

[5] D. E. Atkinson et al., Next-generation performance of SAPHIRA HgCdTe APDs, in High Energy, Optical, and Infrared Detectors for Astronomy VII, 2016, vol. 9915, no. August 2016, p. 99150N.

[6] G. Finger et al., SAPHIRA detector for infrared wavefront sensing, in Adaptive Optics Systems IV, 2014, vol. 9148, no. August 2014, p. 914817.

[7] S. B. Goebel, D. N. B. Hall, I. Pastrana, and S. M. Jacobson, HgCdTe SAPHIRA arrays: individual pixel measurement of charge gain and node capacitance utilizing a stable IR LED, in High Energy, Optical, and Infrared Detectors for Astronomy VIII, 2018, vol. 10709, no. July 2018, p. 37.

[8] A. Vaccarella et al., Cryogenic detector preamplifier developments at the ANU, in High Energy, Optical, and Infrared Detectors for Astronomy VIII, 2018, vol. 10709, no. July 2018, p. 80.